\documentstyle{l-aa} 

.tfm    
\def\lsim{\lower.5ex\hbox{$\; \buildrel < \over \sim \;$}}
\def\gsim{\lower.5ex\hbox{$\; \buildrel > \over \sim \;$}}
\begin{document}
\title {Observational evidence for mass ejection during soft X-ray dips in 
GRS~1915+105}
\author{S. V. Vadawale$^1$, A. R. Rao$^1$, A. Nandi$^2$ and 
S. K. Chakrabarti$^2$ }
\institute { $^1$Tata Institute of Fundamental Research, Homi Bhabha Road,
Mumbai(Bombay) 400 005, India\\
$^2$S.N. Bose National Center for Basic Sciences, Salt Lake, Calcutta 700 091,
India}
\offprints { S.V. Vadawale ({\it santoshv@tifr.res.in})}
\date{Received ; accepted , }
\thesaurus{02.01.2; 02.02.1; 08.23.3; 08.9.2 GRS 1915+105; 13.25.5}
\maketitle
\markboth{Vadawale et al: Evidence for mass ejection in GRS~1915+105}{}


\begin{abstract}
We investigate the connection between the X-ray and radio properties 
of the Galactic microquasar GRS~1915+105, by analyzing the X-ray data
observed with RXTE, during the presence of a huge radio flare ($\sim$450 
mJy). The X-ray lightcurve shows two dips of $\sim$100 second duration.
Detailed time resolved spectral analysis shows the existence of three 
spectral components: a multicolor disk-blackbody, a Comptonized component 
due to hot plasma and a power-law. We find that the Comptonized component 
is very weak during the dip. This is further confirmed by the PHA ratio of 
the raw data and ratio of the lightcurves in different energy bands. These 
results, combined with the fact that the 0.5 -- 10 Hz QPO disappears during 
the dip and that the Comptonized component is responsible for the QPO lead 
to the conclusion that during the dips the matter emitting Comptonized 
spectrum is ejected away. This establishes a direct connection between the 
X-ray and radio properties of the source.


\end{abstract}

\medskip

\section{Introduction}

The Galactic microquasar GRS~1915+105 is a bright X-ray source and it is 
a subject of intense study in all wavelengths, particularly in radio and 
X-ray wavelengths (see Mirabel \& Rodriguez 1999 and references therein).
It has been exhibiting different types of X-ray variability characteristics 
(Morgan, Remillard,  \& Greiner 1997; Muno et al. 1999; Yadav et al. 1999; 
Belloni et al. 2000a). The radio emission from this source also demonstrates 
its chaotic nature by means of time to time huge radio flares (Mirabel \& 
Rodriguez 1994; Fender et al. 1999), long episodes of high/low emissions and 
periodic oscillations (Pooley \& Fender 1997). There were several attempts 
in the past to correlate the radio and X-ray emission characteristics. Pooley 
\& Fender (1997) reported short period radio oscillations coincident with 
X-ray dips. Fender \& Pooley (1998) showed that the IR emission, interpreted 
as the high-energy tail of a synchrotron spectrum, also varies on similar 
time scales. Feroci et al. (1999) reported disappearance of inner accretion
disk during a small radio flare. Thus, so far there are many evidences for 
the morphological correlation between X-ray emission and small radio 
oscillations or flares. However, in the case of huge radio flares, exhibited by
this source from time to time, there is no strong morphological identification 
with detailed X-ray emission characteristics. Fender et al. (1999) suggested 
that the repeated ejections of the inner accretion disk (Belloni et al. 1997) 
might be responsible for such flares. It was pointed out that such oscillations,
having hard dips are not always accompanied by high radio emission (Naik \& Rao
2000; Yadav et al. 1999). This suggests that some other mechanism is 
responsible for such huge radio flares. Recently, Naik \& Rao (2000) made 
a systematic study of the morphology of different types of X-ray emission and 
accompanying radio emission and found an one to one correspondence between 
the soft dips in X-rays (observed during classes $\beta$ and $\theta$) and 
high radio emission. Naik et al. (2001) have suggested that the huge radio 
flares might be produced due to a number of such soft dip events.
 
 In this {\it Letter} we propose an evidence of mass ejection, during the soft
X-ray dips, by performing a detailed time resolved X-ray spectroscopy of the 
RXTE archival data observed simultaneously with a huge radio flare. We identify
three components in the spectrum and show that the Comptonized component 
disappears during the dips. We explain this as ejection of the inner cloud 
and thus establish a direct connection between huge radio flares and X-ray 
emission from this source. 

\section{Analysis and Results}

 GRS 1915+105 exhibits huge radio flares from time to time, the most recent 
of which occurred on 1999 June 8. The PPC detectors of IXAE (Agrawal et al. 
1997) observed the source during the entire episode of this radio flare, 
including the low-hard state of the source just prior to the flare. The IXAE 
observations revealed the presence of regular soft dips in the X-ray 
lightcurve during the radio flare. During these dips the X-ray flux decreases
by a factor of three within $\sim$5 seconds, remains low for $\sim 30 - 60$ 
seconds and then gradually recovers to the maximum (Naik et al. 2001). 
Inspired by this observation, we obtained the RXTE data observed on 1999 
June 8 (ObsID: 40702-01-03-00) to study, in detail, the spectral properties 
of the dips during the radio flare. This is the only pointed RXTE observation 
during this flare (Naik et al. 2001). The PCA (Jahoda et al. 1996) lightcurve 
and hardness ratio for this observation is shown in Figure 1. It shows that 
this observation belongs to class $\theta$ as defined by Belloni et al. 
(2000a). The class $\theta$ shows almost regular soft dips of 40-100 seconds 
duration (defined as state A) and variable low-hard state (defined as state C)
outside the dip.  

\begin{figure}[t]
\vskip6.5cm
\includegraphics{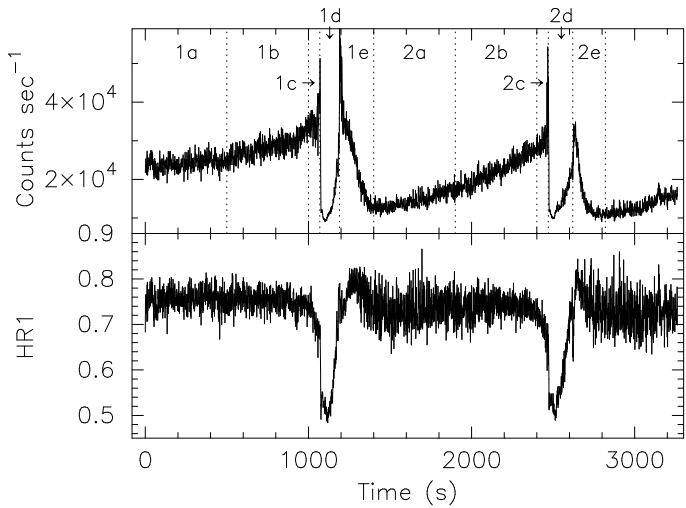}
\caption[]
{
Lightcurve (top panel) and hardness ratio (6-15 keV / 2-6 keV; bottom panel) 
of GRS~1915+105 obtained on 1999 June 8 using {\it RXTE-PCA}. The regions 
chosen for time resolved spectral and temporal studies are shown in the 
top panel, separated by dotted lines.
}
\end{figure}

\begin{figure}[t]
\vskip15.0cm
\includegraphics{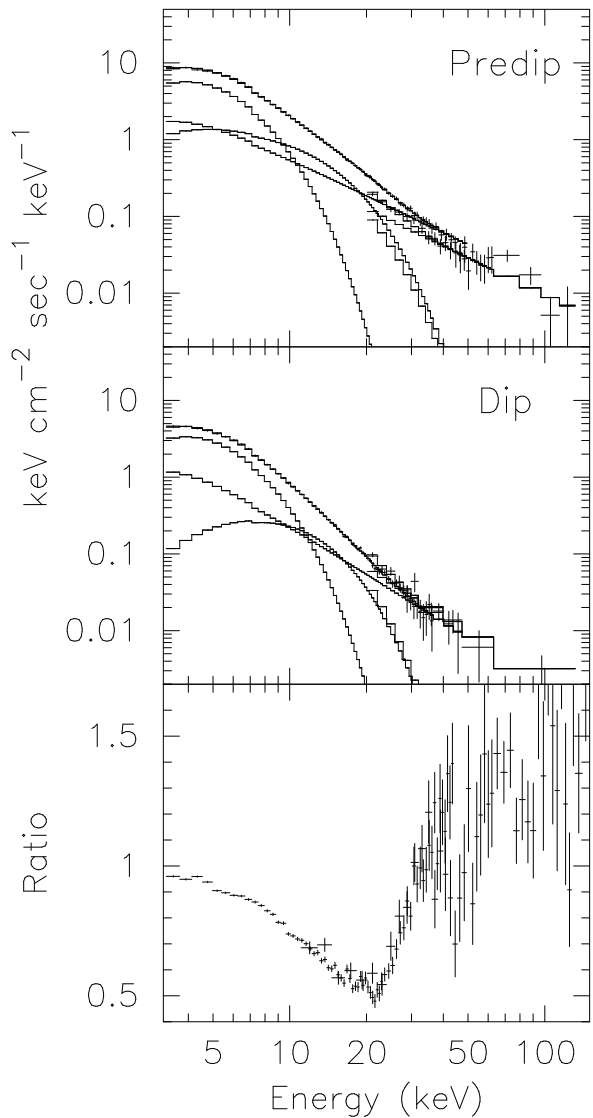}
\caption[]
{
The deconvolved X-ray spectra of GRS~1915+105 during the last 80 seconds 
before the dip (top panel) and 100 seconds during the dip (middle panel). 
The model consists of a disk blackbody, Comptonization due to hot plasma, 
and a power-law. The bottom panel shows the ratio of observed count rates 
(dip to pre-dip ratio) normalized at 30 keV, which highlights the lack of 
counts in the range 8 - 30 keV during the dip. 
}
\end{figure}

\begin{table*}
\caption{Results of temporal and spectral analysis$^1$ of RXTE 
data of GRS~1915+105 during class $\theta$ on 1999 June 8}
\begin{center}
\begin{tabular}{ccccccccccc}
\hline
\hline
Region & $\nu_{QPO}$ & P$_{QPO}$ & ~~$kT_{in}$~~ & ~~$kT_e$~~ & ~~~$\tau$~~~ &
~~~$\Gamma_X$~~~ & \multicolumn{4}{c}{2-50 keV Flux (10$^{-8}$ erg cm$^{-2}$ 
s$^{-1}$) }\\
 & Hz & rms \% & (keV) & (keV) & & & Total & Diskbb & CompST & Power-law\\
\hline
\hline
1a & 5.99 & 3.10 & 1.51 & 3.35 & 16.77 & 2.81 & 6.69 & 2.00 & 1.13 & 3.56 \\
1b & 6.68 & 2.47 & 1.58 & 3.55 & 12.92 & 2.86 & 7.71 & 2.46 & 1.55 & 3.70 \\
1c & 8.34 & 1.89 & 1.51 & 3.18 & 12.85 & 2.61 & 8.80 & 4.25 & 2.20 & 2.34 \\
1d &   -  &  -   & 1.50 & 2.58 & 36.28 & 3.11 & 4.07 & 2.49 & 0.45 & 1.12 \\
1e & 4.67 & 3.44 & 1.56 & 3.52 & 12.86 & 2.75 & 6.27 & 2.12 & 1.28 & 2.86 \\
2a & 4.71 & 6.74 & 1.36 & 3.60 & 13.69 & 2.74 & 4.17 & 0.98 & 0.68 & 2.50 \\
2b & 5.60 & 3.76 & 1.50 & 3.64 & 12.75 & 2.84 & 6.18 & 1.66 & 1.08 & 3.44 \\
2c & 7.27 & 1.65 & 1.45 & 3.14 & 21.53 & 2.60 & 7.52 & 3.66 & 1.90 & 1.96 \\
2d &   -  &  -   & 1.44 & 2.40 & 35.30 & 2.93 & 3.95 & 2.35 & 0.53 & 1.06 \\
2e & 4.20 & 6.57 & 1.42 & 3.52 & 13.66 & 2.61 & 4.44 & 1.42 & 0.96 & 2.06 \\
\hline
\hline
\multicolumn{11}{l}{$^1$The model components are disk blackbody (diskbb), 
thermal-Compton (CompST) and power-law.}\\
\multicolumn{11}{l}{~~~Typical errors  :
~~~Inner disk temperature $kT_{in}:~\pm0.03$;  
~~Temperature of the Compton cloud $kT_e:~\pm0.15$; } \\
\multicolumn{11}{l}{~~~~~~~Optical depth of the Compton 
cloud $\tau:~\pm0.40$;~~Power-law photon index  $\Gamma_X:~\pm0.05$;} \\
\end{tabular}
\end{center}
\end{table*}

 We have attempted a wide band, time-resolved X-ray spectroscopy of the dip 
events by making spectral fits to the data from different portions of the 
lightcurve during both the observed cycles. We have divided each cycle into 
five intervals: pre-pre-dip (a, 500 s), pre-dip (b, 500 s), edge (c, 80 s), 
dip (d, 100 s and 140 s) and post-dip (e, 200 s). Figure 1 (top panel) shows 
the selection of these time intervals. For wide band spectral fitting we have 
extracted 129 channel spectra from PCA and 64 channel spectra from HEXTE. We 
have used data from cluster 0 of HEXTE and have added 2\% systematic error 
to PCA spectra (Vadawale et al. 2001; Gierlinski et al. 1999). The spectra 
during very short intervals e.g. dip and edge, were rebinned to fewer number 
of channels in order to improve the statistics. We have fitted the PCA (3-50 
keV) and HEXTE (15-150 keV) spectra simultaneously with different models 
(see Vadawale et al. 2001 and Rao et al. 2000).

\begin{figure}
\vskip10.0cm
\includegraphics{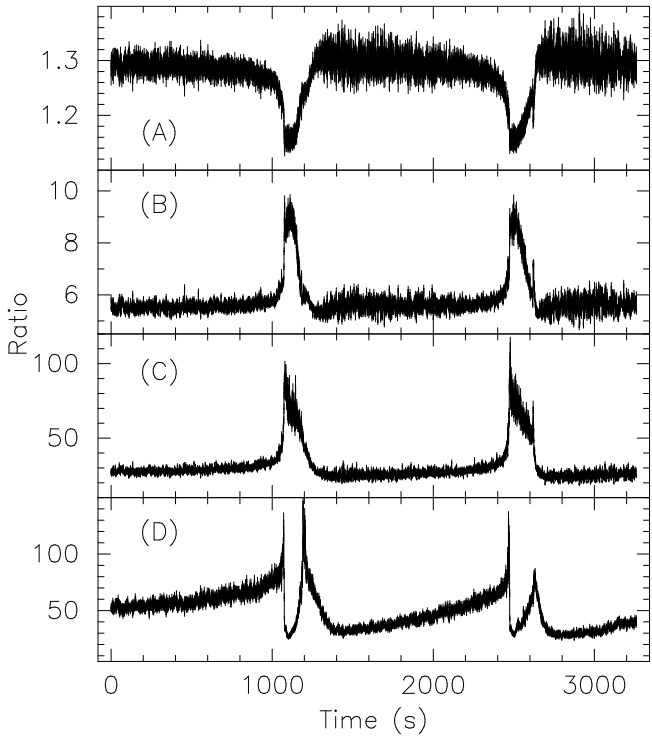}
\caption[]
{
Ratio of the lightcurves of GRS~1915+105 (a) std-1~/~2-8 keV (b) 
std-1~/~8-15 keV (c) std-1~/~15-25 keV (d) std-1~/`25-60 keV. Opposite shape 
of the dip in the top / bottom and middle two panels suggests that during the 
dip, relative decrease of counts in the middle energy ranges is larger than
that in the low / high energy range.
}
\end{figure}

 We find that the X-ray spectrum cannot be fitted by the ``standard'' model 
for the Black Hole Candidates (BHCs), consisting of a disk-blackbody and 
a power-law. It is also known previously that the X-ray spectra of radio loud 
states are peculiar and cannot be described by the standard model (Muno et al.
1999; Belloni et al. 2000b) and hence it is necessary to look for more complex 
models. We find that a three component model consisting of a disk-blackbody, 
a Comptonization due to hot plasma (CompST - see Sunyaev \& Titarchuk 1980) 
and a power-law is necessary for statistically and physically acceptable fit 
to the X-ray spectra of the current observation. For example, in the pre-dip1 
region, a model consisting of a disk-blackbody and a power-law gives 
$\chi^2=272$ for 89 degrees of freedom (dof), a model consisting of a 
disk-blackbody and a CompST gives $\chi^2=191$ for 88 dof, whereas the 
model consisting of three component: a disk-blackbody, a CompST and a power-law
gives $\chi^2=85$ for 86 dof. The spectral fits improve by similar orders in 
the other regions as well, by using the three component model. The same model 
is used by Vadawale et al. (2001) and Rao et al. (2000) to describe the radio 
loud low-hard state of this source, and they give a detailed justification for
the existence of the third component. It should be mentioned here that the wide 
band (3-150 keV) spectral fitting is critical for identification of all the 
three components. Vadawale et al. (2001) have suggested the origin of the 
additional power-law as the very high-energy tail of the synchrotron radiation 
responsible for the radio emission. Our results strengthen their conclusion 
that the high radio emission manifests itself in X-ray spectra as an additional
power-law component. The parameters of the best fit model are shown in Table 1
along with the component-wise X-ray flux for each region separately. The 
observed values of QPO frequency and the rms power in the QPO are also given 
in the table.

The two dip periods are particularly interesting because of the very week 
CompST component and the absence of QPO. The absence of QPO (also reported 
earlier by Muno et al. 1999 and Markwardt et al. 1999), combining with the 
result that only the CompST is responsible for the QPO (Rao et al. 2000), 
suggests that the CompST should also be absent during the dips. The same is 
indicated by the large decrease in the CompST flux compared to other two 
component fluxes (Table 1). A small CompST flux in the dip could be due to the
inclusion of the recovery period in the dip spectrum. It is not possible to get
the combined PCA and  HEXTE-CL0 spectra during the first $\sim$60 second of dip
minima due to the rocking motion of the HEXTE clusters. However, spectral 
analysis of the dip minima using only PCA data shows that a disk-blackbody + 
power-law model gives statistically acceptable fit ($\chi^2=84$ for 60 dof), 
whereas the same model, for the pre-dip1 PCA data, gives unacceptable fit 
($\chi^2=220$ for 64 dof). This leads to a hypothesis that the CompST is 
really absent in the beginning of the dip and slowly reappears during the 
later part of the dip.

To verify this hypothesis, we examined various ratios of the raw data. First 
two panels of Figure 2 show the unfolded spectra of pre-dip and dip intervals,
obtained for the first dip, whereas the bottom panel shows the ratio of the 
observed count rate during the dip period to that during the pre-dip period,
normalized at 30 keV. This ratio clearly shows that the dip period has fewer 
counts in the middle energy range (10-30 keV), in which the spectrum is 
dominated by CompST component, compared to the low and high energy range. 
This justifies our hypothesis that only CompST vanishes during the dip. To 
examine the temporal behavior of the dip in different energy ranges, we show 
in Figure  3, the ratio of the PCA Standard-1 lightcurve (consisting of photons
of all energy) to the lightcurve in different energy ranges. First panel of 
this figure shows that, during the dip, decrease in the count-rate in 2-8 
keV range is less then that in the total count-rate, whereas the second and 
third panel show that, during the dip, decrease in the count-rate in 8-25 keV 
range (where CompST is a dominant component) is more than that in the total
count-rate. Opposite shape of the dip in the first two panels shows that the 
decrease in the count-rate during the dip is strongly energy dependent and 
is most in the range where CompST is dominating. The shape of the dips in the 
fourth panel, which are shallower than the dips in the total lightcurve and 
thus show the effect of the dips in the lightcurve above 25 keV, is also 
opposite to that in the third panel. This provides further evidence to our 
hypothesis by showing that the decrease in the count-rate at high energies, 
is less than that in the middle energies. 

Thus figures 2 \& 3 provide strong support to our hypothesis, made from the 
time resolved spectroscopy, that the CompST component disappears during the 
dips. This can be interpreted as the ejection of the matter of the Compton 
cloud.

\section{Discussion}
Microquasars are thought to be the Galactic analogues of the distant quasars 
and AGNs. Because of their very low mass, compared to the AGNs, they provide 
a unique opportunity to probe the astrophysics of the AGNs in very short time 
scales. GRS~1915+105 is one of the most active microquasars, and shows all 
possible modes of the mass inflow and outflow, exhibited by means of highly 
complex emission throughout the electromagnetic band. Belloni et al. (2000a) 
classified the different types of the X-ray emission from this source in 12 
different classes and the present observation belongs to the class $\theta$, 
which predominantly consists of the low-hard state and soft dips. Our finding, 
that the X-ray spectrum of the low-hard state outside the dips in this 
observation is best described by three component model corroborates the 
conclusion drawn by Vadawale et al. (2001), who found that the X-ray spectra 
of all radio-loud low-hard states require the three component model.

The vanishing of the Comptonized component during the dip leads to the 
interpretation that the matter responsible for the Comptonized component is 
ejected away from the inner region of the accretion disk and the ejected 
matter emits the synchrotron radiation which is observed as the radio flare. 
As time progresses, this matter is replenished and the Comptonized component 
reappears. Nandi et al. (2001) estimate a mass of $\sim$10$^{18}$ g to be 
ejected during a dip event, based on the TCAF model, and they give a physical
basis for such an ejection. The radio flare which occurred on 1999 June 8 is
fairly large, with flux at 2.25 GHz reaching up to 500 mJy, which are 
previously observed only during the superluminal ejection from this source. 
Rodriguez \& Mirabel (1999) have estimated a typical mass of the superluminal 
ejecta of the order of 10$^{22}$ - 10$^{23}$ g and hence a collection of a 
large number of such dips can cause the ejection of the superluminal ejecta.
These results give a concrete support to the suggestion made by Naik et al.
(2001) that the huge radio flares are produced by multiple dip events.

  We wish to point out here that the hard state (state C) outside the soft 
dips, and variations in them, are also associated with radio emission in 
GRS~1915+105. Yadav (2001) has found a correlation between the X-ray hardness 
ratio in state C and the strength of radio emission for various X-ray 
variability classes. Belloni et al. (2000b) have estimated the mass accretion 
rate from the changes in the sizes of the inner accretion disk and have 
associated them with the outflow rates. In this work, we have found a definite 
evidence from X-ray spectroscopic analysis for a particular emission region 
disappearing during the soft dips (state A). It is quite possible that the 
state C (and variations in them) is associated with flat spectrum radio 
emission (see also Vadawale et al. 2001) and the soft dips are associated with 
steep spectrum radio emission coming from the superluminally moving ejecta. 
A continuous X-ray and radio observations during a superluminal jet emission 
episode should throw further light on the origin of radio emission in 
GRS~1915+105.

\section{Conclusion}
 In this {\em Letter} we have shown that during soft X-ray dips, decrease in 
the count rate in the middle energy range (8-25 keV) is significantly greater 
than that in the low / high energies. We explain this as an ejection of a 
Compton cloud, whose radiation dominates in the middle energy range. We suggest
that a collection of a large number of such dips can eject the required mass 
of the superluminally moving blobs.
 
\begin{acknowledgements}

We thank referee for useful comments and suggestions to improve the 
presentation of this paper. 
This research has made use of data obtained through the High Energy
Astrophysics Science Archive Research Center Online Service, provided by the
NASA/Goddard Space Flight Center.

\end{acknowledgements}

{}

\end{document}